\documentclass[aps,prb,twocolumn,groupedaddress]{revtex4}

\usepackage{amsmath}
\usepackage{graphicx}

\begin{document}


\title{Multiscale Modeling of a Nanoelectromechanical Shuttle}

\author{C. Huldt}
\email[]{chuldt@fy.chalmers.se}

\author{J.M. Kinaret}
\affiliation{Department of Applied Physics, Chalmers University of
  Technology, S-412 96 G\"{o}teborg, Sweden}

\date{\today}

\begin{abstract}
In this article, we report a theoretical analysis of a
nanoelectromechanical shuttle based on a multiscale model that
combines microscopic electronic structure data with macroscopic
dynamics. The microscopic part utilizes a (static) density functional
description to obtain the energy levels and orbitals of the shuttling
particle together with the forces acting on the particle. The
macroscopic part combines stochastic charge dynamics that incorporates
the microscopically evaluated tunneling rates with a Newtonian
dynamics.

We have applied the multiscale model to describe the shuttling of a
single copper atom between two gold-like jellium electrodes. We find
that energy spectrum and particle surface interaction greatly
influence shuttling dynamics; in the specific example that we studied
the shuttling is found to involve only charge states $Q=0$ and
$Q=+e$. The system is found to exhibit two quasi-stable shuttling
modes, a fundamental one and an excited one with a larger amplitude of
mechanical motion, with random transitions between them.

\end{abstract}

\pacs{to-be-found}


\maketitle

\section{Introduction}

Nanoelectromechanical systems
that combine electrical and mechanical functionalities on the
nanometer scale have in the recent years attracted a great deal of theoretical
and experimental interest.\cite{Roukes01,Craighead00} The nanoelectromechanical shuttle is
a structure that resembles a single electron transistor but incorporates
mechanical motion of the central island. Previous theoretical works on the shuttle
have shown that
in the presence of an DC applied bias
the charge and velocity of the central island are correlated,
$\overline{Q(t)\dot{Z}(t)} \neq 0$, which implies that 
the shuttle absorbs energy from the DC field and converts it
into mechanical motion. The shuttle motion facilitates charge
transfer through the system, and signatures of mechanical motion
can be seen both in the current-voltage characteristics and in 
the noise properties of the device. \cite{Gorelik98,Shekhter03,Flindt05}

Several theoretical
studies have been carried out for different setups of
the shuttle since the first description of this phenomenon. The
theoretical studies cover different size regimes of the shuttle,
featuring coherent\cite{Fedorets02} or
sequential\cite{Nord02,Braig04} tunneling and quantum
mechanically\cite{Boese01,McCarthy02}
or classically\cite{Isacsson98,Nishiguchi03} described mechanical motion. The studies have shown
that the shuttle instability strongly depends
on the bias voltage and the system setup. This sensitivity
also renders the shuttle behavior
dependent on the precise description of the problem.

Experimental evidence of coupling between vibrational degrees of
freedom and electron transfer has been found for both microscopic\cite{Scheible02,Erbe01}
and macroscopic\cite{Tuominen99} systems. In particular, the experiment
by Park {\it et al.}, reference \onlinecite{Park00}, using a C$_{60}$ molecule between gold electrodes has
demonstrated the type of coupling that has been considered by many theoretical
studies and has increased the interest
for a molecular shuttle.\cite{Boese01,Fedorets02,Nishiguchi03,McCarthy02,Braig03}

In the shuttle geometry the mechanical motion is on a nearly macroscopic
time scale, typically from picoseconds for small molecules
to nanoseconds for large molecules such as carbon nanotubes. The motion
is excited due to tunneling events between the mobile object and the 
stationary electrodes, which have a typical timescale of femtoseconds and
are determined by the electronic structures of the mobile molecule and the 
electrodes. Hence, a theoretical description of the shuttle system naturally
calls for multiscale methods that combine the fast electronic time scales with
the slower mechanical ones. Thus far research has concentrated on the slow
degrees of freedom while dealing with the fast ones in a phenomenological
approximation.

The two main issues addressed in this work are the impact 
of the electronic structure of the central island on the shuttling motion,
and an analysis of the short range interactions between the central island
and the stationary electrodes. The first issue we will address by describing
the central island using density functional theory,\cite{Hohenberg64,Kohn65} which provides information
on the energy spectrum of the island as well as structure of the relevant orbitals.
The interaction between the island and the electrodes is described in part by 
a phenomenological Born-Mayer potential combined with image charge effects,
and in part by a model that transfers mechanical energy from the shuttle to
lattice vibrations in the electrodes, thereby dissipating energy of the shuttle
system and preventing catastrophic runaway. Due to the phenomenological description
of surface interactions, some physical effects such as chemisorption are not 
properly accounted for which limits the applicability of the present model to
materials for which chemisorption can be neglected. A way to overcome
this problem in a future study would be to incorporate a time-dependent DFT\cite{Runge84} 
module that describes both the island an the electrode during the crucial parts
of the shuttle cycle; at present, however, that type of description is prohibitively 
expensive from a computational point of view.

The DFT data is used to evaluate tunneling matrix elements and tunneling rates
between the central island and the electrodes. These are then inserted to a dynamic
module that describes both the charge dynamics in terms of stochastic tunneling events
and mechanical motion using molecular dynamics. The resulting macroscopic dynamics
is implemented as a dynamic Monte Carlo algorithm that uses the output of a series of
static DFT calculations as its input. This stochastic dynamics can be used to address both
average transport properties such as current and random fluctuations, or noise, which 
both exhibit clear signatures of shuttling according to phenomenological theories.

\section{Method}

\subsection{Setup}

\begin{figure}
\includegraphics[width=0.35\textwidth]{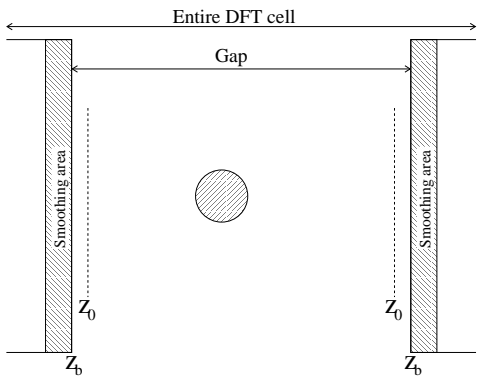}%
\caption{\label{fig:DFTcell} The system. A copper atom is placed in
between two jellium surfaces 15 {\AA} apart. A buffer with the width
of 2.5 {\AA} is added on both side of the gap for the DFT
calculations in order to localize the electrons. A small smoothing
area between gap and barrier is needed to speed up calculations.}
\end{figure}

For computational efficiency, we have chosen to focus on the
simplest possible system where the central island comprises
just one atom. However, the methods and qualitative results should be
applicable also to more complex system. The system we consider
includes two electrodes 15 {\AA} apart, described as
semi-infinite jellium slabs. The
central island is a copper atom that can move in a direction normal to
the electrode surfaces. For the electrodes, the Wigner-Seitz radius is set to
3 a.u. and the electrode work function $W$ is set to 3.5 eV.
\cite{Kiejna} The fermi energy, $\epsilon_{F}$, is
calculated from the Wigner-Seitz radius while other material
specific electrode parameters are taken from gold. A bias voltage of
3 V is applied over the gap, the potential dropping from left to
right. The region described by the DFT module consists of the region
between the two electrodes plus a buffer region of 2.5 {\AA} inside
each of the electrodes as show in Fig. \ref{fig:DFTcell}. The buffer
regions are needed so that the plane-wave-based code can better describe
orbitals that are localized in the inter-electrode gap.

\subsection{Electronic structure calculations}

The electrodes and the space between them are treated separately: the
electronic structure of the shuttling object is described using a density functional
code that is limited to the inter-electrode space while the electrodes
themselves are described analytically within the jellium approximation.
For the electronic structure calculation we use the
DaCapo code\cite{DaCapo} with the PW91 exchange correlation functional\cite{PW91}  and employ
the adiabatic approximation to separate the electronic structure calculation
from the dynamics description.

The effect of the electrodes on the island is
modeled as a one-electron potential and inserted
directly into the effective potential in the Kohn-Sham equations.
The potential is divided into two parts: the interaction
between an electron and a metal surface, and the interaction between
an electron and the induced charge arising from the remaining charges
in the space between the electrodes.
For the former part we use the saturated image barrier
\begin{equation}\label{eq:Jennings}
V_{J}(z)= \left\{\begin{aligned}
-\frac{q^{2}}{16\pi\epsilon_{0}(z-z_{0})} &  (1-e^{-\lambda(z-z{0})})
& \vspace{5mm} z \geq z_{0}\,\\
-\frac{V_{0}}{Ae^{B(z-z_{0})}+1} &  & z \leq z_{0},\end{aligned}
\right.
\end{equation}
suggested by Jones and Jennings.\cite{Jones84} Here the parameters
are related to the work function $W$ and Fermi energy $\epsilon_F$
of the electrodes through $V_{0}=W+\epsilon_{F}$,
$A=-1+16\pi\epsilon_{0}V_{0}/q^{2}\lambda$ and
$B=8\pi\epsilon_{0}V_{0}/q^{2}A$ where the parameters $z_{0}$ and
$\lambda$ are found by fitting eq. (\ref{eq:Jennings}) with DFT
calculated tabulated data on the effective potential, $v_{{\rm
eff}}$.\cite{Lang70} The values were determined to $\lambda=1.6$
\AA$^{-1}$ and $z_{0}=0.34$ {\AA} outside the jellium background
$z_{b}$. The suppression factor $(1-\exp (-\lambda\,(z-z_{0})))$,
which multiplies the classical image potential, tends towards zero
as the mirrored charge closes in on the surface. The image plane,
$z_{0}$, is used as the effective surface of the material, in
compliance with Lang and Kohn.\cite{Lang73} The value of $z_{0}$
derived for eq. (\ref{eq:Jennings}) differs from the value derived
by Lang and Kohn for {\it e.g.} image charge
potential.\cite{Jennings88PRB} However, for the interactions with
the induced charge, the individual contributions from the electrons
and the nucleus should cancel out far from the surface and it is
therefore suitable to keep only one parameter for the effective
surface.

For the second term, the potential at ${\bf r}$ due to a charge at position
${\bf r}'$, we use a form $v({\bf r};{\bf r}')$
that satisfies Poisson's equation
in the region outside the electrodes and reduces to the saturated image potential
if the mirrored charge is at the same point where the potential is
measured giving
\begin{equation}
v(\textbf{r};\textbf{r}^{\prime})=-\frac{q^{2}}{4\pi\epsilon_{0}\mid
\textbf{r}-{\bf \tilde{r}}^{\prime} \mid}\left( 1-e^{-\lambda
(z^{\prime}-z_{0})}\right)
\end{equation}
The resulting potential for interaction with the core is
\begin{equation}\label{eq:Onepotnucl}
V_{In}({\bf
  r})=Q_{v}\,v(\textbf{r};\textbf{Z}),
\end{equation}
where $Q_v$ is the core charge of the island, {\em i.e.}, the number of protons minus the number
of core electrons.

The model for the interaction with the induced charge from the other
electrons is obtained similarly. Here we assume that the electron
distribution of the island is well approximated with a Gaussian, as
is typically the case, and integrate over the mirrored charge to
obtain
\begin{equation}\label{eq:Onepotel}
V_{Ie}({\bf
  r})=-(Q_{e}-1)\int d{\bf r^{\prime}}v(\textbf{r};\textbf{r}^{\prime})\rho({\bf
  r}^{\prime}).
\end{equation}
where
\begin{equation}
\rho({\bf r}^{\prime})=\frac{1}{(2\pi)^{3/2}\sigma^{3}}e^{-\frac{1}{2}({\bf
Z}-{\bf
  r^{\prime}})^2/\sigma^2}
\end{equation}
is a form function and $Q_{e}$ is the number of valence electrons
for the central island. The width $\sigma(Q)$ is calculated by
fitting a Gaussian to the unperturbed electron density for different
charge states, $Q$ denoting the number of extra charge units on the
central island. Assuming localization of electrons to the island or
the leads (no chemisorption), $Q$ is strictly integer. 

The added bias voltage is
\begin{equation}
V_{bias}(z)=E(z-\Delta_{gap}/2)
\end{equation}
where the electric field is $E=V/(\Delta_{gap})$ and $\Delta_{gap}$
the distance between the electrodes. Finally, the Pauli repulsion that confines electrons within the gap
is for the electronic structure calculations described as a
repulsive square potential wall placed at $z_{b}$. The repulsion
term is important for the separation of electronic structure
calculation and charge transfer mediated by tunneling, however, the
form of the repulsion is less important as most of the tunneling
events take place when the island is relatively far from the surface
on an atomic scale.

Owing to the time-consuming and time-independent character of DFT,
it is not possible to determine the system properties continuously.
Instead, the simulations are performed on a number of positions and
charge states. A continuous description is produced
with interpolation.

\subsection{Forces}

The forces on the atomic core are calculated
using the DFT code. In
the Kohn-Sham single-particle equations all charges but one electron are treated as
a mean-field static charge distribution. We implement the potential due to
surface interactions as an external field. This results in proper (mean field) orbitals and
energy eigenvalues but incorrect forces and total energies.
Therefore, a correction term in the form of $\frac{\partial
V_{DFT}}{\partial Z}-\frac{\partial V_{el}}{\partial Z}$ is added to
the forces calculated  by the electronic structure code. Here, $V_{DFT}$ is the induced
charge potential used in the DFT calculation, the sum of eq. (\ref{eq:Jennings}),
(\ref{eq:Onepotnucl}) and (\ref{eq:Onepotel}). The actual
potential, $V_{el}$ is obtained as a variational derivative of the full electronic
energy as
\begin{equation}
\begin{aligned}
V_{el} = & \frac{Q_{e}}{2}\int d^{3}{\bf
 r}^{\prime}\left[v(\textbf{Z};\textbf{r}^{\prime})+v(\textbf{r}^{\prime};\textbf{Z})\right]\rho({\bf
 r}^{\prime})\\
 & -Q_{v}v(\textbf{Z};\textbf{Z}) + g({\bf r})\end{aligned}
\end{equation}
where the last term $g({\bf r})$ is independent of $Z$ and does not contribute to the
forces.

A smooth many-body short-range repulsion is added to the total
force. We have chosen a Born-Mayer type pair
potential\cite{Gilbert68} and integrated over the electrode surface
which yields the force
\begin{equation}
F(z)= f_{0}\,e^{(\sigma_{Au}+\sigma_{Cu})/\rho)} \left(e^{z/\rho}+2\pi
  z \rho n_{s} e^{(-\sqrt{z^{2}+a^{2}/(2\pi)}/\rho}\right).
\end{equation}
For a Cu-atom outside a gold surface, the effective radii are
$\sigma_{Cu}=0.77$ {\AA} and $\sigma_{Au}=1.37$ {\AA},
$n_{s}=2/a^{2}$ is the surface density of atoms and $a_{lat}=4.078$
{\AA} is the lattice constant for a fcc [100] gold
surface.\cite{Shannon76,Ph.Hb} The closest electrode lattice site is
considered separately (in order to maximize the localization) while
the other lattice sites are handled as a continuum. The values
$f_{0}$ and $\rho$ are 6.95 10$^{-11}$N and 0.3 {\AA}
respectively.\cite{Kunz92,Baxter03,Varsamis01,Adams01}

\subsection{Transition rates}

Charge transfer rates between the central island and electrodes are
calculated with the transfer Hamiltonian method\cite{Bardeen61,Duke}
using the overlap between the atomic and electrode wave functions. The transition
rates to and from a specific atomic orbital are

\begin{equation}\label{eq:Gamma1}
\Gamma_{\rightarrow p}=\frac{V\,n_{f}}{(2\pi)^{3}}\int\! d^{3}k\, \frac{2\pi}{\hbar} \mid
M_{pq}\mid^{2}\delta\left(-W_{R/L}+\frac{\hbar^{2}k^{2}}{2m}-E_{p} \right)
\end{equation}
where
\begin{equation}\label{eq:RewriteBardeen}
M_{pq}=\int_{z \in R_{gap}}\! d^{3}r\, \psi^{*}_{p,a}(V_{a}-V_{m})\psi_{q,m}.
\end{equation}
Here, $\psi_{p,a}$ is a Kohn-Sham orbital (localized on the
island) with eigenenergy
$E_{p}$, $\psi_{q,m}$ is a metal wave function and $n_{f}$ is the
relevant number of states available for tunneling in the (often degenerate)
orbital $\psi_{p,a}$. The
potentials $V_{a}$ and $V_{m}$ are the effective potentials for
$\psi_{p,a}$ and $\psi_{q,m}$ respectively. The atomic orbital
$\psi_{p,a}$ and the effective gap potential $V_{a}$ are extracted
from the DFT simulations. The transition rates are calculated
numerically for all energetically allowed transitions determined by comparing the
electrode chemical potentials with the Kohn-Sham eigenvalues.

The electrode wave functions are calculated analytically from a
square potential with a finite barrier at the electrode surface and
a hard wall at $-\infty$. The electrode wave functions in the
$z$-direction are
\begin{equation}\label{eq:psizmetal}
\psi_{k_{z}}(z)= \left\{ \begin{aligned}
\sqrt{\frac{2\kappa_{z}}{\kappa_{z}L+1}}\sin \delta_{z}e^{-\kappa_{z}z}\hspace{5mm} & \hspace{10mm} z \geq 0\\
\sqrt{\frac{2\kappa_{z}}{\kappa_{z}L+1}}\sin(k_{z}z-\delta_{z}) &
\hspace{5mm}-L \geq z \geq 0\end{aligned} \right.
\end{equation}
with $\kappa_{z}=\sqrt{\frac{2mW}{\hbar^{2}}-k_{z}^{2}}$ and
$\delta_{z}=\arcsin(\frac{\hbar k_{z}}{\sqrt{2mW}})$. A small offset between the physical and the geometrical
surface is implemented to attain charge neutrality of the jellium
slabs.\cite{Kiejna} Periodic boundary
conditions are assumed parallel to the surface.

\subsection{Dynamics}

A dynamic Monte Carlo approach is used to calculate the shuttle
dynamics. Input parameters to this module consist of core forces and
transition rates as functions of the island position. The motion of the
central island is described classically by
\begin{equation}\label{eq:NII}
m\ddot{x}=F_{ext}(x,Q)+F_{dissip}(x,Q)
\end{equation}
where $F_{ext}$ are the core forces given by the previous
calculations and $F_{dissip}$ is a dissipation term. The island
position $x(t)$ is calculated by numerically integrating the
equation of motion, while the island charge $Q(t)$ is allowed to
change stochastically using the tunnel rates determined above. This
results in a coupled stochastic dynamics for the mechanical and
electrical degrees of freedom.

As the shuttle absorbs energy from the bias voltage,
the dissipation term is essential for the stability of the island
motion.\cite{Gorelik98,Fedorets04} Earlier theoretical work has mainly used viscous damping,
$-\eta \dot{x}$.\cite{Gorelik98,Nishiguchi03,Smirnov04,Chtchelkatchev04,Pistolesi05} In this work a different model based on mechanical
damping is used, coupling a simple model of the surface to the
island equation of motion via the surface forces as
\begin{equation}\label{eq:eqmotion}
\left\{\begin{array}{lcl}
m\ddot{x}&=&F_{tot}(x)-X\,F^{\prime}_{tot}(x)\Theta(X\,F^{\prime}_{tot}(x)\dot{x})
\\
M\ddot{X}&=&-\dot{M}\dot{X}-kX-F_{tot}(x-X), \end{array}\right.
\end{equation}
where $x$ and $m$ are the position and mass of the central island,
$X$ and $M$ are the position and mass of the surface, and $k$ is an effective
surface spring constant. The $\Theta$-function in the first equation
restricts the energy flow so that the shuttle energy can be transferred to
lattice vibrations of the electrodes (phonon emission) while the opposite
process of phonon absorption is forbidden.

The equations have been formulated in terms of the surface element nearest
to the shuttling object, which implies that the effective mass of the surface
depends on the separation between the mobile island and the surface; this can
be determined by assuming an elastic model for the surface and requiring that (i) the
instantaneous displacement  $X(t)$ agrees with that of the surface atom nearest
to the shuttling object,  and (ii) the total momentum of the surface is $M\dot{X}$.
The elastic parameters for surface atoms have been chosen to correspond to those
of gold, and effective parameters in the simplified model are consequently
$M(X) = m_{Au}F(X)/F_{max}(X)$, where $F(X)$ is the total force between the shuttling
object and the surface and $F_{max}(X)$ is the force on the surface atom
nearest to the shuttle and $k = k_{Au}F(X)/F_{max}(X)$ where $k_{Au} = 8m_{Au}v_s^2/a^2$
is the spring constant obtained from the sound velocity $v_s$ and lattice constant $a$.

The main qualitative difference between our dissipation model and
viscous damping is that in our model dissipation occurs primarily
when the island-electrode interaction is strongest. This influences
the threshold voltage for onset of shuttling, and also renders the threshold
dependent on the initial conditions.

\section{Results}
 
\begin{figure}
\includegraphics[width=0.45\textwidth]{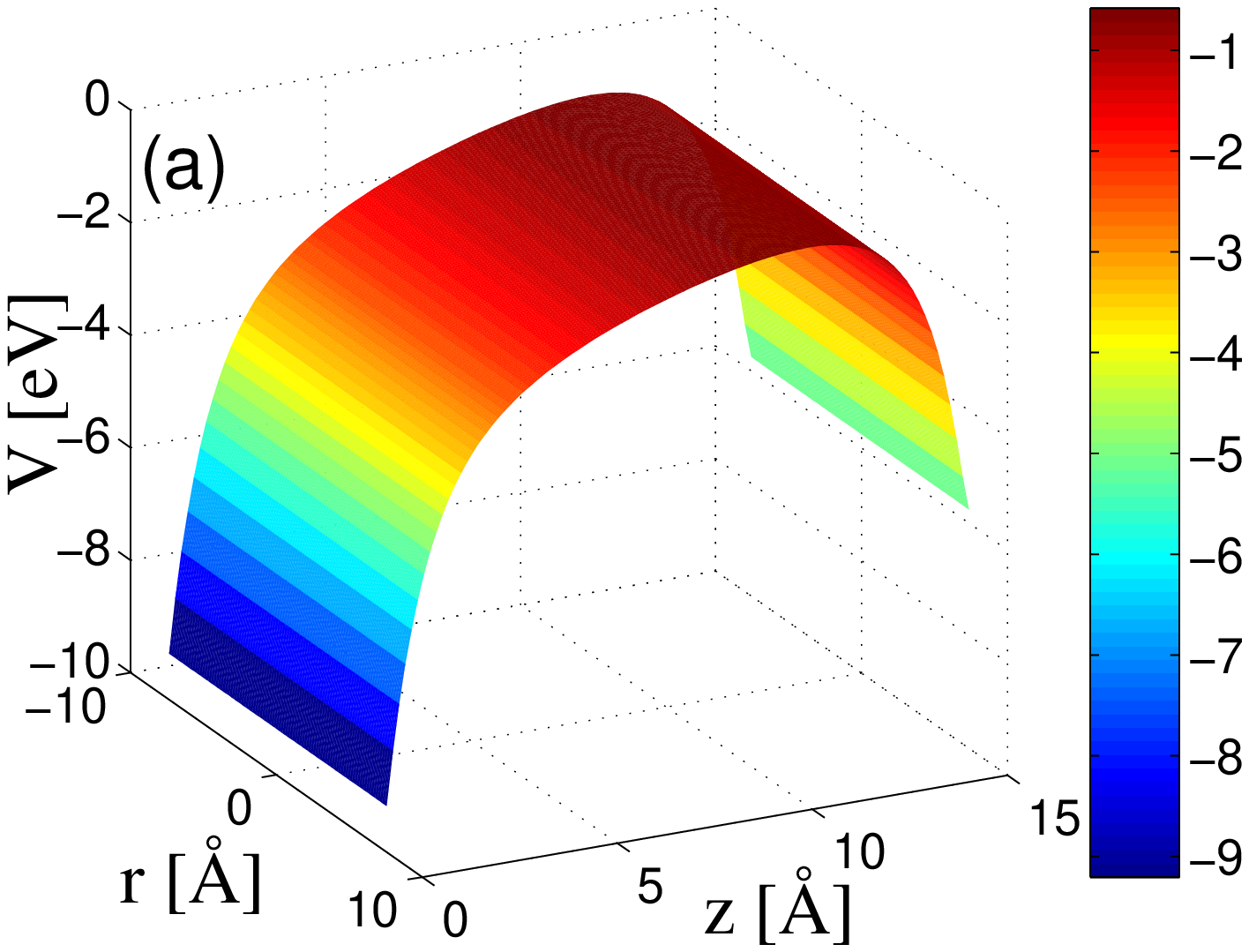}\\
\includegraphics[width=0.45\textwidth]{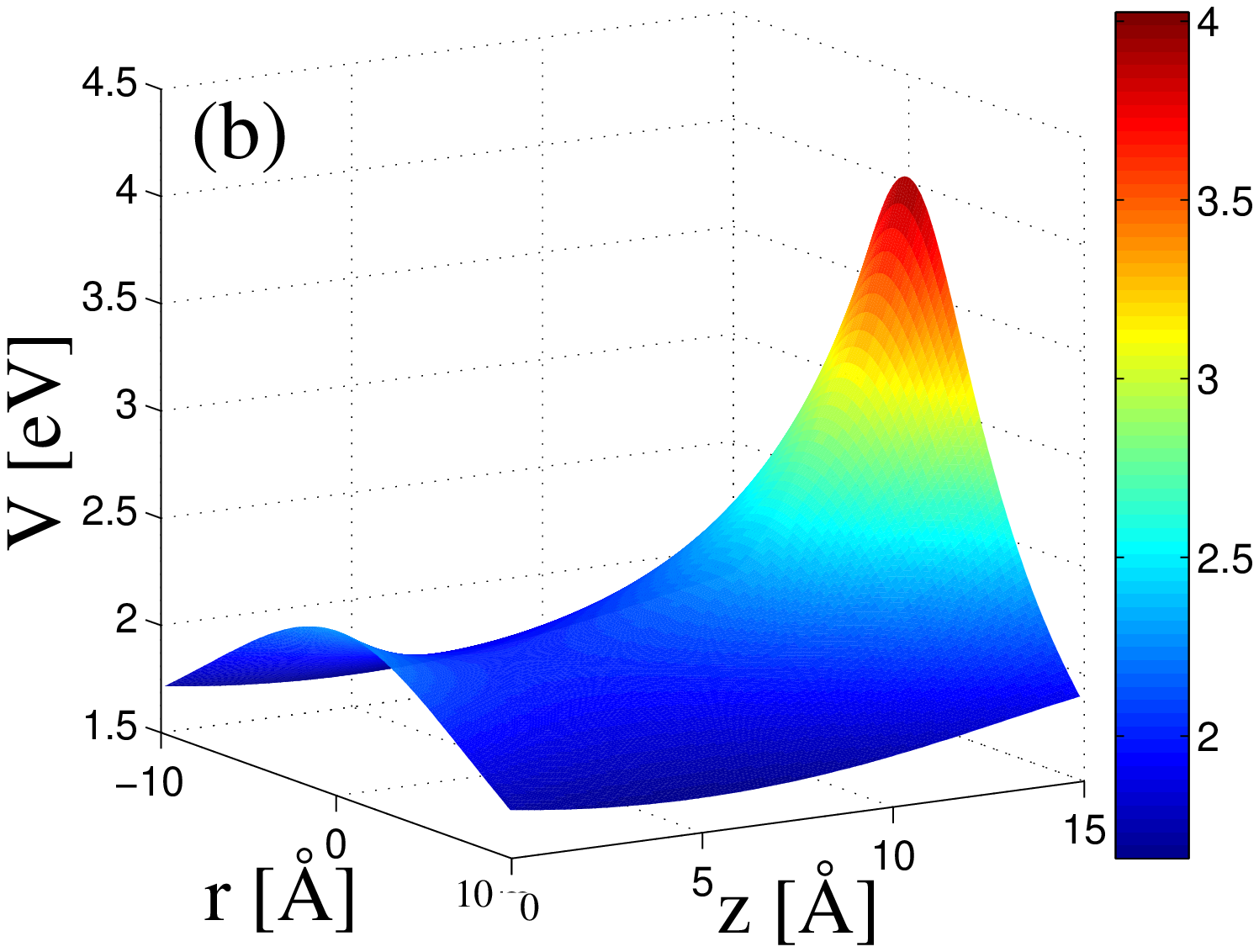}\\
\includegraphics[width=0.45\textwidth]{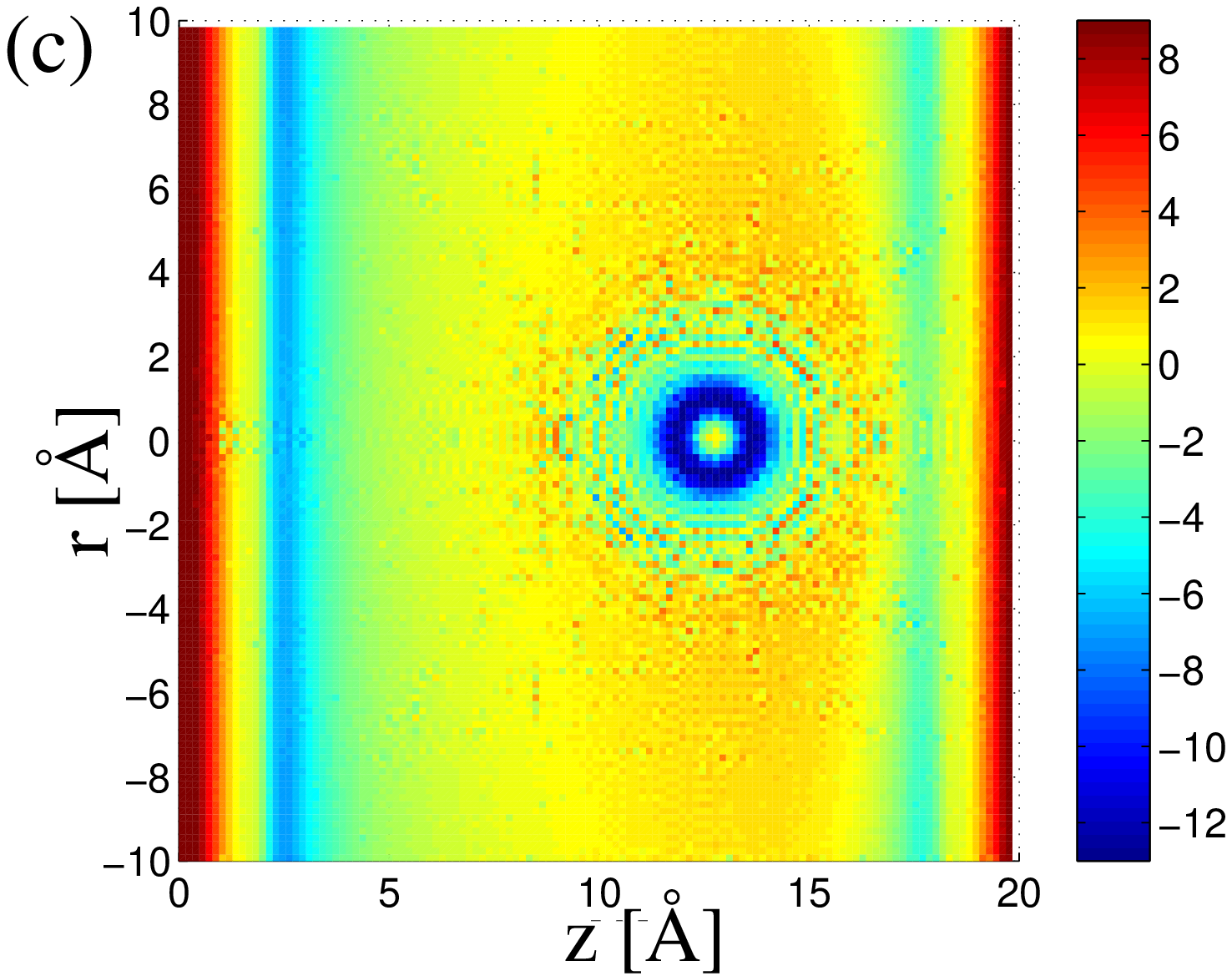}%
\caption{\label{Fig:pot} The Kohn-Sham one-electron effective
  potential comprises several parts rendering diverse behaviors for
  different positions and charge states. The above figures are for
  $Q=0$ and $Z=10.17$ {\AA}. The $z$-axis is the direction of
  island motion, $r$ is parallel to the electrodes. In (a) and (b) $z$
  encompasses the gap while (c) includes the entire DFT cell with the
  2.5 {\AA} buffer regions on both sides of the gap. (a)
  Equation (\ref{eq:Jennings}) and the bias voltage.
 (b) One electron interaction with induced charge due to other system
  charges. (c) The effective potential as used by
  DaCapo. Close to the electrodes eq. (\ref{eq:Jennings}) forms deep wells.}
\end{figure}

The different parts of the one-electron potential are depicted in
Fig. (\ref{Fig:pot}). The small widths of the form function ($\sigma
\;\sim$ 0.24-0.27 {\AA} for an unperturbed pseudopotential,
$Q=1\ldots -1$, and $\sim$ 0.30-0.35 {\AA} for the double junction
potential) imply that sufficiently far from the surface the point
charge approximation would be quite accurate: for distances $\geq$ 2
{\AA} from the surface, the spatial distribution of charge has
little effect on the potential. For island positions close to the
electrode surfaces, the spread in the valence electron distribution
and the rapid saturation effectively bares the core image making the
effective potential strongly repellent.

\begin{figure}
\includegraphics[width=0.45\textwidth]{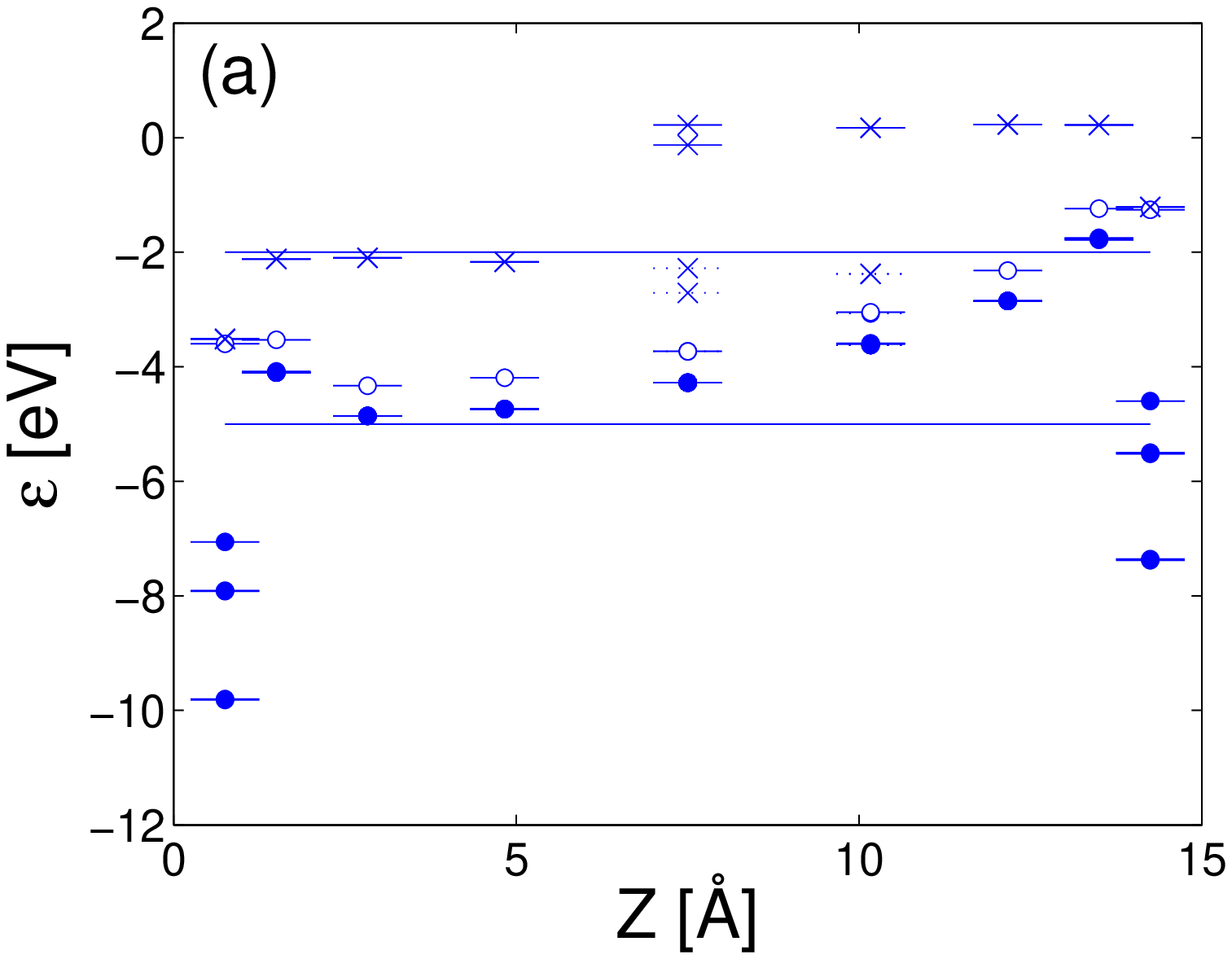}\\
\includegraphics[width=0.45\textwidth]{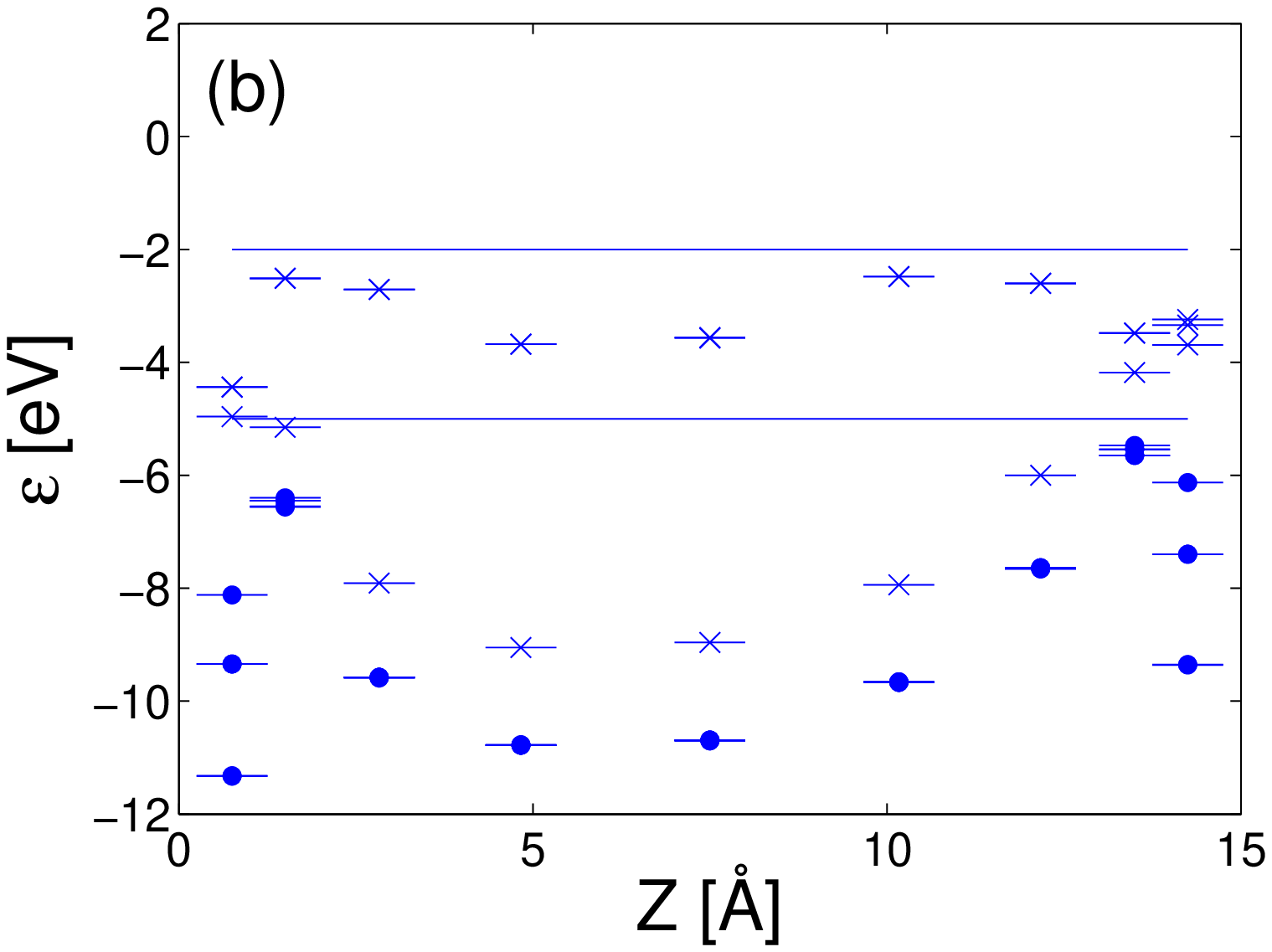}\\
\caption{\label{Fig:eig}The lowest eight non-spin polarized
Kohn-Sham eigenvalues for nine positions of the central island.
Levels indicated by filled circles are
  fully occupied, empty circles half-occupied and crosses unoccupied.
  The upper and lower solid lines are
  $\mu_{R}$ and $\mu_{L}$ respectively. (a) $Q=0$.
  The lower lying excitations (core positions in the left half of the gap) have
  been calculated including the surface potential well near the electrode.
  The higher lying excitations (core positions near right electrode) have been
  calculated without the surface potential well near the left electrode as transitions
directly from the right electrode to the left surface well have very small tunneling rates.
For $Z=7.5$ {\AA} and $Z=10.17$ {\AA} results are
  depicted for both approaches (solid and dashed lines). For the
  positions closest to the electrodes, the width of the
  surface potential wells causes a substantial drop in eigenenergies. (b)
  $Q=1$.}
\end{figure}

The resulting Kohn-Sham eigenvalues, depicted in Fig. \ref{Fig:eig}, are
used as energy spectra for the central island. Comparison between
the eigenvalues an the electrode chemical potentials gives the
possible transitions. Full relaxation into the $N/2$ lowest bands
is assumed instantaneous, where $N$ is the number of valence
electrons (11 for the used Cu GGA pseudopotential, $Q=0$). Higher bands
are treated as excitations.\cite{Baerends97} The temperature is taken to be zero in the treatment
of tunneling events; however, in the DFT calculation a finite temperature is
needed for convergence.

A small correction is needed for some calculations in order to use
the equilibrium DFT calculations within a dynamic picture. For some
island positions and $Q \leq 0$, it is energetically favorable to
place some of the extra charge in the surface potential well outside the
positive electrode surface instead of on the central island.
However, the time scale for this direct equilibration between leads
is very long. In order to find the energy spectra and orbitals that
are relevant for the dynamic evolution, the surface well near the left
electrode surface is manually suppressed for core positions near the
right electrode; due to the polarity of the applied bias, similar
problems do not arise for core positions near the left electrode.
The possibility of transitions directly
between the electrodes is kept, but the transition rates are small
enough to be of no importance for the results.

For the tunneling rate calculation the Kohn-Sham eigenvalues are
regarded as electron energies, which is known to be rigorously
correct for the ionization potential involving the HOMO
level,\cite{Baerends97} and believed to be reasonably accurate for
the other levels as well.\cite{Chong02} We assume that tunneling
rates are sufficiently low so that the island fully relaxes between
each tunneling event to the configuration determined by
time-independent DFT. The time scale for this relaxation is
typically in the femtosecond range which is fast compared to the
tunneling rates except for core positions very close to the
electrode surfaces; however, since energetics severely limits the
possible tunneling processes, the instantaneous relaxation
approximation is reasonably well justified for all core positions.
For the chosen bias voltage, the possible charge transitions for the
island are $1 \rightarrow 0$ and $0 \rightarrow 1$.

It is interesting to notice the asymmetry of attainable charge
states that arises from the asymmetry of the energy spectrum of the
Cu atom: the dynamical evolution only involves charge states $Q=0$
and $Q=1$ but not $Q=-1$. The symmetric expression for charging
energy $E=Q^{2}/2C$ used for larger metallic grains is only
justified if the level spacing on the island is small enough so that
the electrostatic energy scales dominate. This asymmetry implies
that the shuttling is asymmetric also in the sense that energy is
absorbed from the DC field only during half a period which makes the
system more sensitive to dissipative mechanisms.

The occupied Kohn-Sham eigenfunctions are
identified as $d$- and $s$-orbitals in accordance with the expected electron
configuration of Cu, 3$d^{10}$4$s^{1}$. Close to the electrode surfaces, the
orbitals deform against the repulsion wall. For all but the
closest position to the electrodes the 4$s$-orbital gives the widest
electron distribution and the largest contribution to the transition
rates.

\begin{figure}
\includegraphics[width=0.45\textwidth]{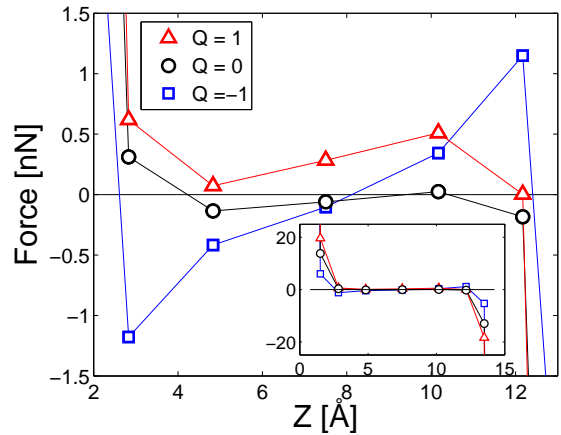}%
\caption{\label{Fig:Force} Total forces on the grain core, center
  positions. Positions not included $Z=0.75$ {\AA} and $Z=14.25$ {\AA} reach $\pm
  (500-650)$ nN. Lines between positions correspond to the used
  interpolation scheme.}
\end{figure}

The core forces for the central positions are strongly
dependent on the delocalization of the electron distribution of the central
island (Fig. \ref{Fig:Force}). For the positive ion with $Q=1$ the dominant force is the
electrical bias while the potential of $Q=-1$ is a nearly
symmetric image charge potential. For $Q=0$, the sign depends
on the description of the surface interactions, and with the interaction model
we have chosen the neutral atom feels a slight net force towards the negatively charged
right electrode. The repulsion from the surface is dominant
for the two outermost positions on either side giving a physisorption
minimum between 3--5 {\AA} from the electrode surface.

\begin{figure}
\includegraphics[width=0.45\textwidth]{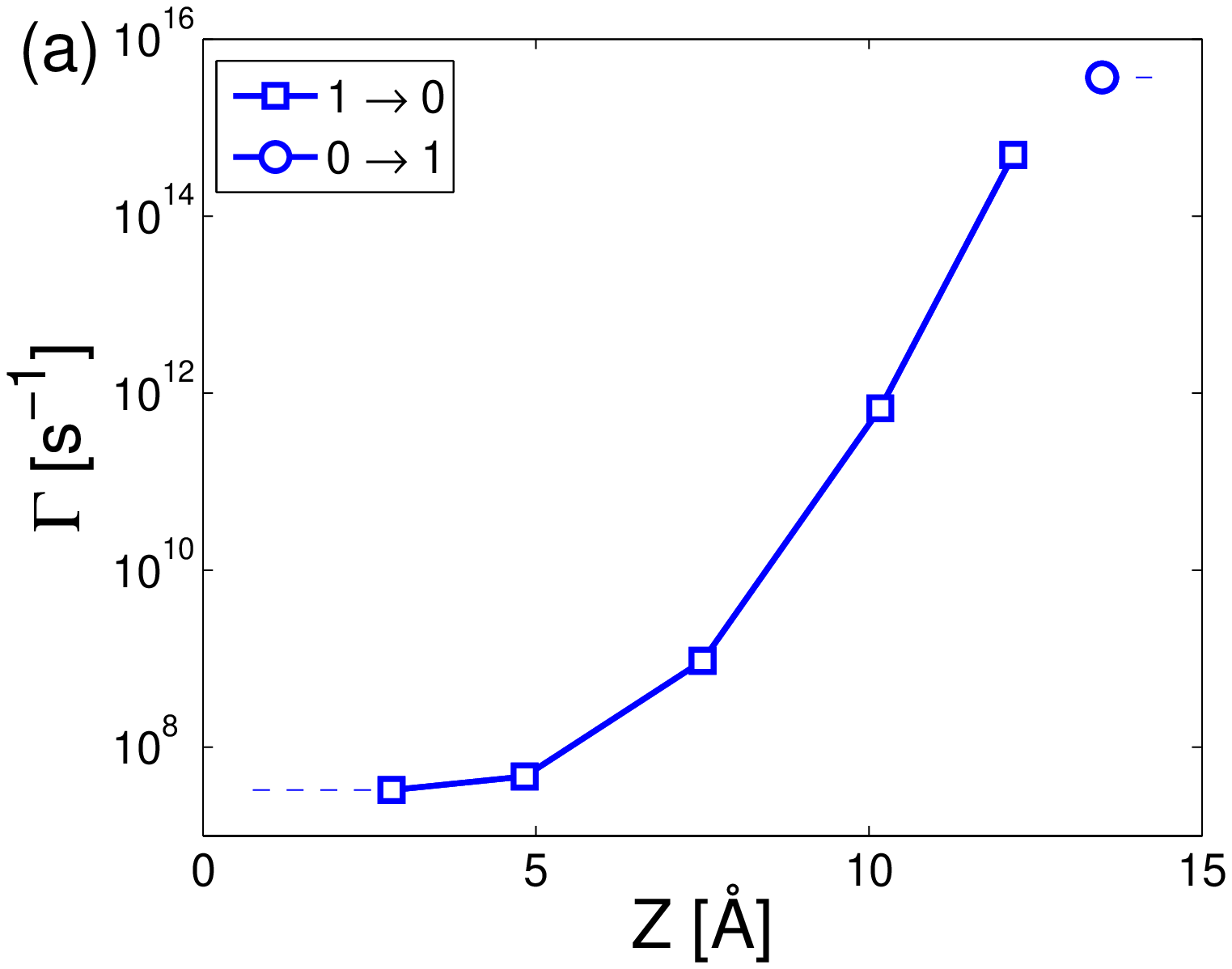}\\
\includegraphics[width=0.45\textwidth]{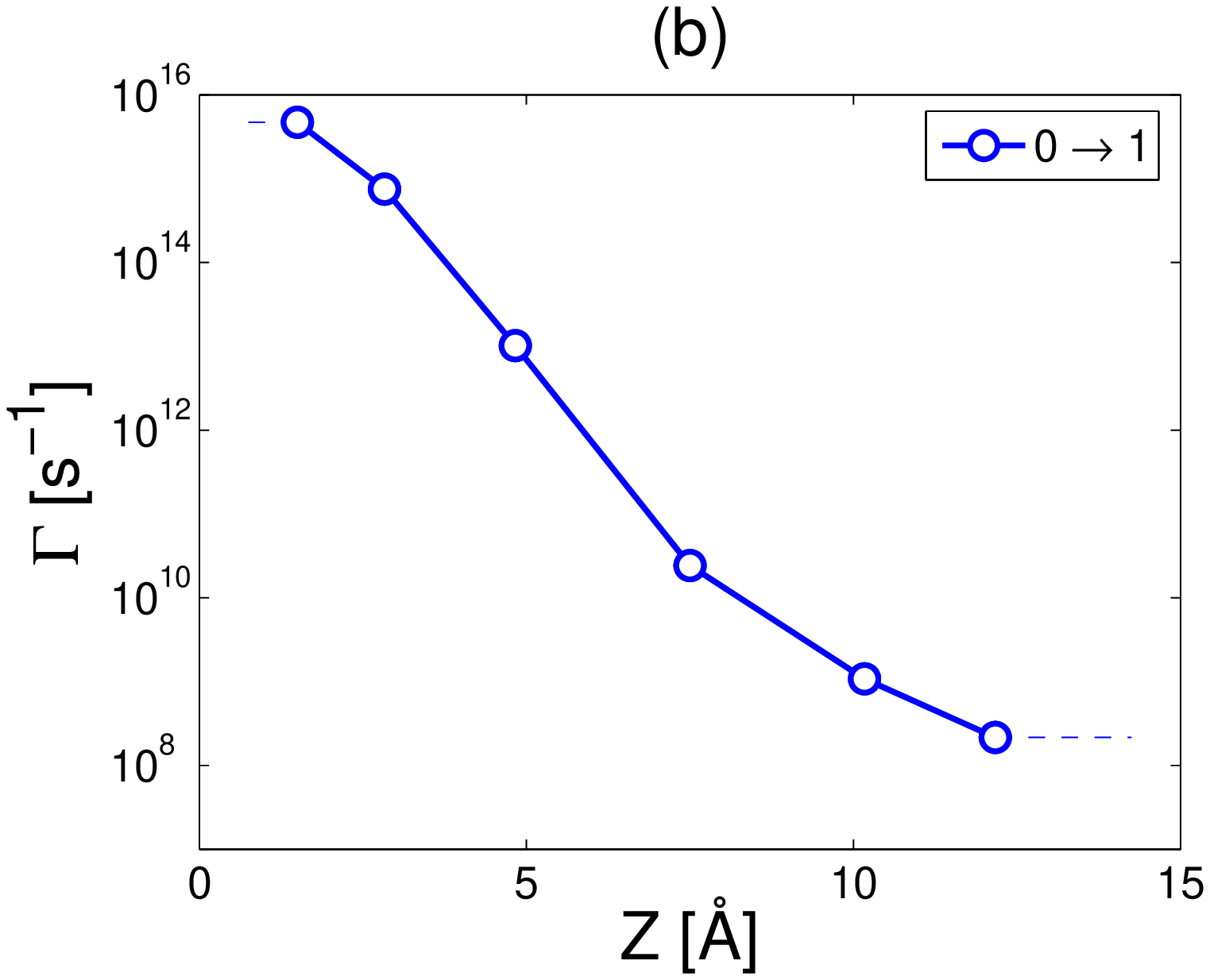}%
\caption{\label{Fig:Trans}Transition rates: electron current from
  right to left. (a) Squares: Transitions from negative right lead to island.
Close to the negative lead, current mediating
transfer is blocked by energetics (Coulomb blockade). Instead, an electron can
transfer against the bias back to the negative lead, (circles). (b) Electron
transitions from island to positive left lead.}
\end{figure}

The transition rates are much less sensitive to the surface
description than the forces (Fig. \ref{Fig:Trans}). Their distance
dependence is approximately exponential as assumed by effective
theories\cite{Kulik75} with slight saturation for core positions
nearest to the electrodes with a tunneling length that is
approximately 0.4 {\AA} with some variation for the different
allowed transitions. Near the electrodes the energetics
considerations inhibit tunneling, as seen in Fig. \ref{Fig:eig},
which can be viewed as a molecular equivalent of Coulomb blockade.

\begin{figure}
\includegraphics[width=0.45\textwidth]{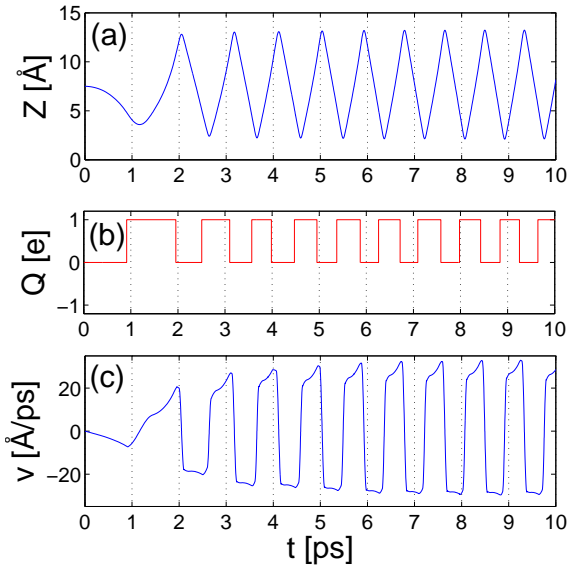}%
\caption{\label{fig:shuttling} Shuttle regime for initial conditions
  $Z(0)=0$, $Q(0)=0$ and $v(0)=0$. (a) Island position. (b) Charge
  state as a function of time. The shuttle carries one electron per
period. (c) Island velocity. The main acceleration and deceleration are close to the
electrodes.}
\end{figure}

In the dynamics simulations, the calculated forces and transition
rates are joined. The result is indeed a stable shuttling regime where $\overline{Q(t)\dot{Z}(t)} \neq 0$.
We have performed dynamical simulations starting from a variety
of initial states, and seen that for most starting conditions the
results are quite similar: as a rule, the model does indeed shuttle electrons (Fig.
\ref{fig:shuttling}). However, for some initial configurations
such as $Z(t=0) \approx 10$ {\AA} and $Q(t=0) = 0$ the applied bias
of 3 V is not sufficient to initiate shuttling.
 One of the more prominent differences between our results and
previous works is the complexity of the forces, particularly near the electrode
surfaces. Since the surface description we use is adapted from static
analyzes, it is unclear how accurately it captures the
interactions in a dynamic situation, and the detailed results are somewhat
uncertain. A slightly different potential renders the forces on the
neutral atom positive over a larger range of positions, and the
range of initial conditions that would result in shuttling would be
smaller, implying that the threshold voltage for shuttling depends sensitively
on the model for surface-island interactions and on the initial conditions.

The detailed structure of the forces of the middle
positions is less important after shuttling is well established.
The main forces become the
close-range exponential forces of the electrodes and the applied
electric field. For the asymmetric shuttle, energy is absorbed by the
charged shuttle from the
field during half a cycle while during the other half-cycle, after an
elastic collision with the electron surface, a neutral
shuttle moves nearly freely in the opposite direction. The energy loss
during the shuttle-electrode collision cannot exceed the energy absorbed
from the field if a stable periodic motion is to be established.

\begin{figure}
\includegraphics[width=0.45\textwidth]{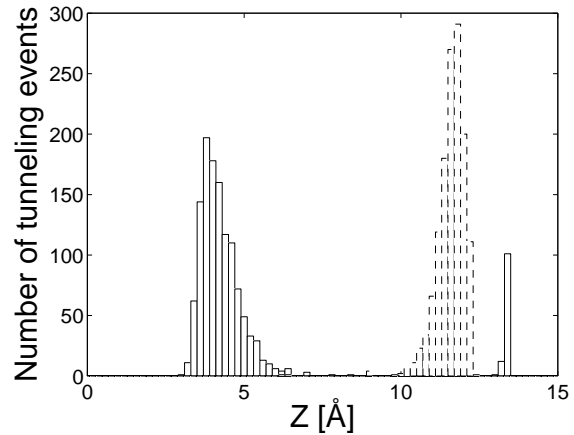}%
\caption{\label{fig:location} The statistical distribution of charge
  transfer locations. The solid and dashed line corresponds to $0 \;
  \rightarrow \; 1$ and $1 \; \rightarrow \; 0$ transitions
  respectively. The quicker growing probability of a $1 \; \rightarrow
  \; 0$ transition compresses the distribution of event locations.}
\end{figure}

The distribution of positions at which tunneling
events take place depends on the transition rates, and the width of the
distribution is connected to the spatial derivatives of the
rates, {\em i.e.} on the tunneling lengths.
The steeper $1 \; \rightarrow \; 0$ gives a more compact
distribution as seen in Fig. $\ref{fig:location}$.

\begin{figure}
\includegraphics[width=0.45\textwidth]{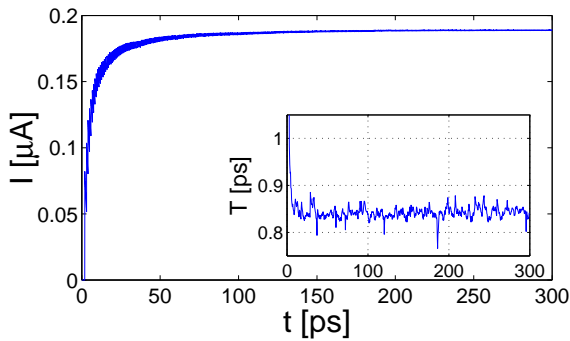}%
\caption{\label{fig:current} The average current, $n_{R}/t$, where $n_{R}$ is the number of electrons from
  the negative lead. Inset: The shuttle
  period as a function of time. The shuttle quickly reaches a stable motion.}
\end{figure}

\begin{figure}
\includegraphics[width=0.45\textwidth]{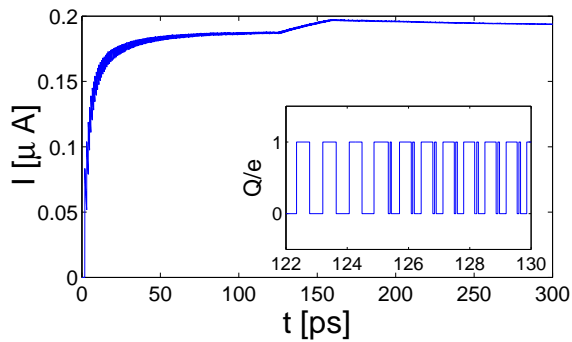}%
\caption{\label{fig:Iquirk} The average current for a path with
  system excitation due to back-tunneling
  close to the negative lead (see inset). The increase of amplitude caused by the larger $Q=1$
forces enhances the possibility for the tunneling triplet to take
place also in the next period. The transition is sharp with an
amplitude increase of $\sim$ 0.3 {\AA} and a current of $\sim$ 0.23
$\mu$A. The excitation lasts until a period without reverse tunneling
takes place.}
\end{figure}

The shuttle reaches a stable shuttling motion quickly with a current
of $\sim$0.19 $\mu$A and an amplitude of $\sim$11.1 {\AA}, (see Fig.
\ref{fig:current}). The random character of the transition processes
influences the turning points very little. There is, however, a
possibility for the system to undergo semi-stable excitations due to
randomness of transfer events and the position dependence of the
energy spectrum near the right lead (Fig. \ref{fig:Iquirk}). Very
close to the negative (right) lead, there is a possibility of a
process in which an electron first tunnels from the electrode to an
initially positively charged ($Q=1$) shuttle that continues to move
towards the right lead, followed by tunneling against the bias back
into the lead, and finally a new tunneling event after the shuttle
has changed its direction of motion. During the time that the
charged shuttle spends near the electrode surface after the second
tunneling event, it experiences a larger force than a neutral
shuttle would, which allows it to absorb more energy from the
potential and results in an enhanced shuttling amplitude. The
increase in amplitude enhances the possibility for this sequence of
three tunneling events to take place also in the next period. The
excitation lasts until a transfer without reverse tunneling takes
place near the negative lead. For the system we have studied, the
amplitude of this excited cycle is about 0.3 {\AA} larger than that
of the simple cycle, and the current level is increased by
approximately 20\% to 0.23 $\mu$A. The possibility of two stable
shuttling amplitudes has recently been discussed by Usmani and
co-workers. \cite{Usmani}

\section{Discussion}

Both the energy spectra and the (Kohn-Sham) orbitals of small molecules
near metal surfaces exhibit a rich structure and vary substantially as
a function of the molecule-metal surface separation. The
transition rates are largely exponential functions of the tunneling
distance as assumed in phenomenological
theories, but the allowed transitions are determined by the energy
spectrum, and in particular near the surfaces certain transitions are
forbidden by energy considerations. This results in
an asymmetry in possible charge states and in asymmetric shuttling
where energy is absorbed only during one half cycle of the periodic motion.

The forces in the system are highly sensitive to the description of
the electrode-molecule interaction and to the electronic structure of
the shuttling object. In the stable shuttling regime the island
velocity is large enough for the island to bounce between the
repulsion walls, and the details of the forces near the middle of
the gap are less important than the balance between dissipation and
short range surface forces. The short range forces are hard to
describe quantitatively due to the increased importance of many-body
effects, deformation of molecular orbitals near surfaces and the
details of the dynamic charge transfer processes. However, for large
enough speed on impact the mobile molecule may bounce off the
surface and establish stable periodic motion.

The shuttle excitations depicted in Fig. \ref{fig:Iquirk} are an
example of effects that arise due to the details of the energy spectrum
of a small system. For a more complicated spectrum and larger bias
voltage more phenomena of the same type can be expected; for a slightly
different model of shuttle-surface interactions we have even observed
that the regular shuttling motion may pass into a more chaotic
behavior. Therefore, it is likely that a microscopic
picture of both forces and energy levels is paramount for both
quantitative and qualitative predictions of molecule-sized shuttles.

\begin{acknowledgments}
We acknowledge fruitful discussions with Robert Shekhter, Tomas Nord, and 
Elsebeth Schr\"oder. This work was supported in part by the 
European Community FP6
    funding through the CANEL project, contract
      no. FP6-2004-IST-003673. This publication reflects the views of the
    authors and not necessarily those of the EC. The Community is not
    liable for any use that may be made of the information contained
    herein

\end{acknowledgments}

\bibliography{Refs}

\end{document}